\documentstyle[epsf,seceq,twoside]{ptptex}
\notypesetlogo  
\title{Property of Chiral Scalar and Axial-Vector Mesons\\
 in Heavy-Light Quark System}
\author{%
Muneyuki {\sc Ishida} and Shin {\sc Ishida}$^*$
}
\inst{%
Department of Physics, Tokyo Institute of Technology\\
Tokyo 152-8551, Japan\\
$^*$ Atomic Energy Research Institute, 
College of Science and Technology\\
Nihon University, Tokyo 101-0062, Japan
}
\abst{%
We investigate the properties of new scalar $X_B,X_D$ and axial-vector 
$X_{B^*},X_{D^*}$ mesons, 
which are recently predicted, in the covariant meson-classification scheme,
to exist as the partners 
of ground-state pseudoscalar $B,D$ and vector $B^*,D^*$ mesons, 
respectively, realizing a 
linear representation of chiral symmetry 
for the light quark component in heavy-light-quark meson system.
The mass splittings between the respective chiral partners are 
shown to be equal,
and the decay widths of one pion emission of $X_B$, $X_D$, $X_{B^*}$ 
and $X_{D^*}$ become to have a common value, 
due to chiral and heavy quark symmetry.   
}

\begin{document}
\maketitle

\setcounter{tocdepth}{4}

\section{Introduction}

The property of heavy-light quark meson systems is one of the important topics
of high energy physics.
Until recently the mass spectra and weak transitions 
of $B$ and $D$ meson systems have been analyzed mainly by using the  
heavy quark effective theory (HQET).\cite{rfHQET} 
The HQET is based on heavy quark symmetry
(HQS), which represent a spin-flavor symmetry of QCD 
in the heavy quark sector.

The QCD has another symmetry, chiral symmetry, in the light quark sector.
Thus far for many years, the chiral symmetry had been applied in hadron physics mainly 
to describe the properties of 
${\mib \pi}$-meson octet through non-linear representation. 
Recently the existence of light $\sigma$-meson,
 $\sigma$(500--600), 
is suggested strongly by the several groups,\cite{rfsigma} 
including ours.\cite{rf555} 
The observed properties of $\sigma$-meson
are consistent\cite{rfmw} with the relation between mass and width 
predicted by the linear 
$\sigma$ model (L$\sigma$M),\cite{rflsm} and this fact seems to suggest that 
the linear representation of chiral symmetry 
is also realized\cite{rflr} in nature. 

In a previous paper\cite{rfCLC} 
we have proposed a new covariant framework for meson level-classification 
and considered chiral transformation for the
$q\bar q$ wave function (WF) directly in the framework of 
covariant oscillator quark model (COQM).\cite{rfCOQM}
The $\sigma$ is argued to be discriminated from the ($q\bar q$) 
$^3P_0$-state appearing in the non-relativistic quark model (NRQM),
 and to be classified into, so to speak, the ``relativistic $S$-wave" 
state of the ($q\bar q$) system, which has never been appeared in NRQM. 
This type of new covariant state is called ``chiralons."
We assume the chiral symmetry for light quark component in the heavy-light 
quark systems is also effective, where 
the possible existence of this new type of mesons 
is suggested as chiral partners of the ground $S$-wave state pseudoscalar
and vector mesons, respectively. 

In this paper we investigate the properties of these new scalar 
and axial-vector mesons by taking into account 
HQS for the heavy quark 
component, as well as chiral symmetry for the light quark component. 
Similar approach was given\cite{rfWise} from a viewpoint of the 
non-linear 
chiral Lagrangian, including only the ground-state psedoscalar 
and vector heavy-light quark mesons and the $\pi$-meson octet.
In this work we adopt the linear representation, and 
make a chiral Lagrangian including also the new scalar and 
axial-vector mesons in the heavy-light $B,D$ systems.

\section{Constraints on One Pion Emission Process from HQS}
For $n\bar b=(u\bar b,d\bar b)$ systems 
the new scalar and axial-vector mesons 
are denoted as $X_B=\ ^t(X_{B^+},X_{B^0})$ and 
$X_{B^*}=\ ^t(X_{B^{*+}},X_{B^{*0}})$.
They are supposed to be the chiral partners of the pseudoscalar 
$B=\ ^t(B^+,B^0)$ and vector $B^*=\ ^t(B^{*+},B^{*0})$, respectively.
Similarly, for $n\bar c=\ ^t(u\bar c, d\bar c)$ system, 
the scalar $X_{\bar D}=\ ^t(X_{\bar D^0},X_{D^-})$ and 
the axial-vector $X_{\bar D^*}=\ ^t(X_{\bar D^{*0}},X_{D^{*-}})$
are supposed to be the chiral partners of pseudoscalar 
$\bar D=\ ^t(\bar D^0,D^-)$ and vector 
$\bar D^*=\ ^t(\bar D^{*0}, D^{*-})$, respectively.

We consider general constraints from HQS on the processes of 
one pion emission,
$X_B\to B\pi$, $X_{B^*}\to B^*\pi$, $X_D\to D\pi$ and $X_{D^*}\to D^*\pi$,
which are expected to be the main decay modes of the relevant mesons. 

The HQ spin symmetry relates $B$ and $X_B$, respectively, to 
$B^*$ and $X_{B^*}$, 
and $D$ and $X_D$, respectively, to $D^*$ and $X_{D^*}$,
with the same velocity. 
Thus, the $S$-matrix elements of the relevant decay
modes satisfy the relations 
\begin{eqnarray}
\langle \pi B({\mib 0}) | U_I | X_B({\mib 0}) \rangle &=&  -i
\langle \pi B^*({\mib 0},\epsilon^{(0)}) | U_I | X_{B^*}({\mib 0},\epsilon^{(0)}) \rangle ,\nonumber\\
\langle \pi D({\mib 0}) | U_I | X_D({\mib 0}) \rangle &=&  -i
\langle \pi D^*({\mib 0},\epsilon^{(0)}) | U_I | X_{D^*}({\mib 0},\epsilon^{(0)}) \rangle ,
\label{eqhqs}
\end{eqnarray}
where $U_I$ is translational operator of the time 
from $-\infty$ to $+\infty$, and ${\mib 0}$ means 
the three velocity ${\mib v}={\mib 0}$.
The longitudinally polarized states of $B^*$, $D^*$ and $X_{B^*}$, 
$X_{D^*}$ appear in RHS. When the $z$-direction is taken as a quantization axis, 
the above equations are derived by using the relation,  $[U_I,S_Q^z]=0$, 
$(2S_Q^z)^2=1$, 
$2S_Q^z| X_B({\mib 0})\rangle= | X_{B^*}({\mib 0},\epsilon^{(0)}) \rangle $ 
and $\langle B({\mib 0}) |(2S_Q^z)= -i 
\langle B^*({\mib 0},\epsilon^{(0)}) |$ etc.,       
where $S_Q^z$ is the $z$-component of the heavy quark spin operator. 
(The factor $-i$ comes from the convention of the spinor WF\cite{rfCLC}
given in the appendix.) 
We assume that the mass difference between the chiral partners,
$\Delta m_M=m_{X_M}-m_M$ $(M=B,D,B^*,D^*)$,
is sufficiently small compared to $m_M$, that is,
$\Delta m_M \ll m_M \stackrel{<}{\scriptscriptstyle \sim} m_{X_M}$,
but is larger than $m_\pi$. 
In this case, 
when we take the rest frame of initial $X_M$, 
the final heavy-light quark meson $M$ also stays almost at rest, 
although the pion is emitted relativistically with energy $\Delta m_M$.
Thus, in both sides of Eq.(\ref{eqhqs}) the pion momentum $p_{\pi\mu}$ 
takes the same value,
and the initial and final heavy-light quark mesons are at rest. 

The HQ flavor symmetry relates the $B$, $B^*$, $X_B$ and $X_{B^*}$ mesons, 
respectively,
to the $D$, $D^*$, $X_D$ and $X_{D^*}$ mesons with the same velocity.
The HQ $\bar b \to \bar c$ flavor transformation leads to  
the following relations
\begin{eqnarray}
\langle \pi B({\mib 0}) | U_I | 
X_B({\mib 0}) \rangle = 
\langle \pi D({\mib 0}) | U_I | 
X_D({\mib 0}) \rangle  & \equiv &
 -\xi \frac{1}{(2\pi )^3 2} \sqrt{\frac{1}{(2\pi )^3 2 E_\pi } }
\nonumber\\
\langle \pi B^*({\mib 0},\epsilon^{(0)}) | U_I | 
X_{B^*}({\mib 0},\epsilon^{(0)}) \rangle  &=&
\langle \pi D^*({\mib 0},\epsilon^{(0)}) | U_I | X_{D^*}({\mib 0},
\epsilon^{(0)}) \rangle  \nonumber\\
 &\equiv & -i \xi ' 
 \frac{1}{(2\pi )^3 2} \sqrt{\frac{1}{(2\pi )^3 2 E_\pi } } ,\ \ \ \ \ 
\label{eqhqf}
\end{eqnarray}
where a common factor 
$\frac{1}{(2\pi )^3 2} \sqrt{\frac{1}{(2\pi )^3 2 E_\pi } }$
is introduced,\footnote{ 
Here we use the ordinary normalization of the state 
$|B({\mib 0})\rangle \equiv 
a_{B({\mib 0})}^\dagger |0\rangle$ etc., 
where $[ a_{B({\mib p})}, a_{B({\mib p}')}^\dagger ]
=\delta^{(3)}({\mib p}-{\mib p}') $.
Through heavy quark $\bar b \to \bar c$  transformation
the state $|B({\mib 0})\rangle$ is transformed into  
$| \bar D({\mib 0})\rangle $,
since one $B$ meson is changed into one $\bar D$ meson. 
The amplitudes $\xi$ and $\xi '$, defined in Eq.(\ref{eqhqf}), 
are related with the relevant Lorentz-invariant amplitudes,
denoted respectively as $g_{X_B\to B}$, $g_{X_D\to D}$, $g_{X_{B^*}\to B^*}$ and 
$g_{X_{D^*}\to D^*}$, through the formula
 $g_{X_B\to B}=\sqrt{m_{X_B}m_B}\xi$,   $g_{X_D\to D}=\sqrt{m_{X_D}m_D}\xi$, 
 $g_{X_{B^*}\to B^*}=\sqrt{m_{X_{B^*}}m_{B^*}}\xi '$ and   
$g_{X_{D^*}\to D^*}=\sqrt{m_{X_{D^*}}m_{D^*}}\xi '$.
 }
and the momentum conservation factor 
$i(2\pi )^4 \delta^{(4)}(P_{X_M}-P_M-p_\pi )$
is omitted in RHS.
Due to Eq.(\ref{eqhqs}), 
the $\xi$ and $\xi '$ satisfy 
\begin{eqnarray} 
 \xi =\xi ' .
\label{eqxi}
\end{eqnarray}
Thus, all the relevant decay matrix elements are reduced to 
one universal amplitude $\xi$ due to HQS. 

The decay widths of $X_M=X_B,X_{B^*},X_D,X_{D^*}$ are commonly 
calculated as
\begin{eqnarray}
\Gamma_{X_M} &=&
 3 \frac{1}{2m_{X_M}} \frac{|{\mib p}|}{4\pi m_{X_M}}
 (\xi \sqrt{m_{X_M}m_M})^2 
 = 3 \frac{\sqrt{(\Delta m_M)^2 -m_\pi^2}}{8\pi} \xi^2 ,
\label{eqhqrel} 
\end{eqnarray}
where we use the approximation $m_{X_M}\approx m_M$, 
$|{\mib p}|=\sqrt{(\Delta m_M)^2-m_\pi^2}$ is the  
pion CM momentum and 
the factor 3 comes from the isospin freedom of final $|\pi M \rangle$ state.
Due to Eq.(\ref{eqhqrel}) the decay widths of the relevant 
processes are only dependent upon the correponding mass difference
$\Delta m_M$.

\section{Linear Chiral Lagrangian for $B$ and $D$ Systems}

We can simply make a linear chiral Lagrangian for the ground state 
$B$ and $D$ meson systems, if we know their  
transformation law. 
We shall use the following rules, which 
are derived by using the covariant spinor WF,\cite{rfCLC}
as explained in the appexdix. 

The chiral transformation for the light $u,d$ quark fields is given by
\begin{eqnarray} 
n=\left( \begin{array}{c} u\\ d \end{array}\right) 
& \to &   g_V \left( \begin{array}{c} u\\ d \end{array}\right) ,\ \ 
 g_A \left( \begin{array}{c} u\\ d \end{array}\right) ,
\end{eqnarray}
where $g_V={\rm exp}\{ i\frac{{\mib \alpha}\cdot{\mib \tau}+\alpha_0\tau^0}{2} \}$ and  
$g_A={\rm exp}\{ i\frac{{\mib \beta}\cdot{\mib \tau}+\beta_0\tau^0}{2}\gamma_5 \} $. 
Correspondingly, the chiral transformation for the light-heavy $B$ and $X_B$
meson fields is given by
\begin{eqnarray}
(X_B+i\gamma_5 B) & \to & g_V (X_B+i\gamma_5 B),\ g_A (X_B+i\gamma_5 B),
\nonumber\\
(X_{B^*}+ \gamma_5 B^*) & \to & g_V (X_{B^*}+\gamma_5 B^* ),
\ g_A (X_{B^*}+\gamma_5 B^*),
\end{eqnarray}
which is also written in the form
\begin{eqnarray}
(X_B-i\gamma_5 B) & \to & g_V (X_B-i\gamma_5 B),\ g_A^\dagger (X_B-i\gamma_5 B),
\nonumber\\
(X_{B^*}- \gamma_5 B^*) & \to & g_V (X_{B^*}-\gamma_5 B^* ),
\ g_A^\dagger (X_{B^*}-\gamma_5 B^*).
\end{eqnarray}
The chiral transformation for the charge conjugate
$\bar B=(B^-,\bar B^0)$ and $X_{\bar B}=(X_{B^-}, X_{\bar B})$ 
meson fields is consistently given by
\begin{eqnarray}
(X_{\bar B}+\bar B i\gamma_5) & \to & (X_{\bar B}+\bar B i\gamma_5) g_V^\dagger ,\ 
(X_{\bar B}+\bar B i\gamma_5) g_A,\nonumber\\
(X_{\bar B}-\bar B i\gamma_5) & \to & (X_{\bar B}-\bar B i\gamma_5) g_V^\dagger ,\ 
(X_{\bar B}-\bar B i\gamma_5) g_A^\dagger .\\
(X_{\bar B^*}-\bar B^* \gamma_5) & \to & (X_{\bar B^*}-\bar B^* \gamma_5) g_V^\dagger ,\ 
(X_{\bar B^*}-\bar B^* \gamma_5) g_A,\nonumber\\
(X_{\bar B^*}+\bar B^* \gamma_5) & \to & (X_{\bar B^*}+\bar B^* \gamma_5) g_V^\dagger ,\ 
(X_{\bar B^*}+\bar B^* \gamma_5) g_A^\dagger .
\end{eqnarray} 
The chiral transformation rule of 
$\sigma$ and ${\mib \pi}$ fields is given by 
\begin{eqnarray}
(\sigma +i\gamma_5{\mib \tau}\cdot {\mib \pi} ) & \to & 
g_V (\sigma +i\gamma_5{\mib \tau}\cdot {\mib \pi} ) g_V^\dagger ,\ 
g_A (\sigma +i\gamma_5{\mib \tau}\cdot {\mib \pi} ) g_A  ,\\
(\sigma -i\gamma_5{\mib \tau}\cdot {\mib \pi} ) & \to & 
g_V (\sigma -i\gamma_5{\mib \tau}\cdot {\mib \pi} ) g_V^\dagger ,\ 
g_A^\dagger (\sigma -i\gamma_5{\mib \tau}\cdot {\mib \pi} ) g_A^\dagger .
\end{eqnarray}
Then a chiral symmetric Lagrangian for $B$ and $X_B$ mesons is given by
\begin{eqnarray}
{\cal L}_B &=& {\cal L}_B^{\rm free} + {\cal L}_B^{\rm Yukawa},
\end{eqnarray}
where the free part ${\cal L}_B^{\rm free} $ and the Yukawa coupling part
${\cal L}_B^{\rm Yukawa}$ are given,\footnote{
The other chiral symmetric combinations, 
$\langle (\partial_\mu X_{\bar B}+\partial_\mu \bar B i\gamma_5)  
(\partial_\mu X_B-i\gamma_5 \partial_\mu B)  \rangle$,
$\langle (X_{\bar B}+\bar B i\gamma_5)(X_B-i\gamma_5 B) \rangle$ 
and
$\langle (X_{\bar B}+\bar B i\gamma_5) (\sigma 
-i\gamma_5{\mib \tau}\cdot {\mib \pi} ) 
(X_B+i\gamma_5 B) \rangle $ give the same contributions as  
the corresponding terms in Eq.(\ref{eqLagB}), and thus omitted. 
} respectively, by
\begin{eqnarray}
{\cal L}_B^{\rm free} &=&   -\frac{1}{4} \langle   
(\partial_\mu X_{\bar B}-\partial_\mu \bar B i\gamma_5)  
(\partial_\mu X_B+i\gamma_5 \partial_\mu B) \rangle \nonumber\\
&& -  \frac{m_{B0}^2}{4} \langle   
(X_{\bar B}-\bar B i\gamma_5)(X_B+i\gamma_5 B) \rangle ,\nonumber\\
{\cal L}_B^{\rm Yukawa} &=& -\frac{g_B}{4} \langle   
(X_{\bar B}-\bar B i\gamma_5) (\sigma +i\gamma_5{\mib \tau}\cdot {\mib \pi} ) 
(X_B-i\gamma_5 B) \rangle ,
\label{eqLagB}
\end{eqnarray}
the notation $\langle\ \ \rangle$ denoting to take trace 
for spinor and light quark 
flavor indices. The $B$-meson and $X_B$ meson have a common 
bare-mass $m_{B0}$ due to chiral symmetry. 

For the vector $B_\mu^*$ and axial-vector $X_{B_\mu^*}$ meson systems 
the chiral symmetric Lagrangian is given\footnote{
The other chiral symmetric combinations, 
$\langle   
(F_{\mu\nu}^{X_{\bar B^*}}+F_{\mu\nu}^{\bar B^*}\gamma_5)
(F_{\mu\nu}^{X_{B^*}}+\gamma_5 F_{\mu\nu}^{B^*}) \rangle$,
 $\langle (X_{\bar B^*\mu}+\bar B^*_\mu\gamma_5)
(X_{B^*\mu}+\gamma_5 B^*_\mu) \rangle$ and 
 $\langle (X_{\bar B^*\mu}-\bar B^*_\mu \gamma_5)
(\sigma -i\gamma_5{\mib \tau}\cdot {\mib \pi} ) 
(X_{B^*\mu}+\gamma_5 B^*_\mu) \rangle$,
give the same contributions as  
the corresponding terms in Eq.(\ref{eqLagB*}), and thus omitted. 
} 
by
\begin{eqnarray}
{\cal L}_{B^*} &=& {\cal L}_{B^*}^{\rm free} + 
{\cal L}_{B^*}^{\rm Yukawa},\nonumber\\
{\cal L}_{B^*}^{\rm free} &=&   -\frac{1}{8} \langle   
(F_{\mu\nu}^{X_{\bar B^*}}-F_{\mu\nu}^{\bar B^*}\gamma_5)
(F_{\mu\nu}^{X_{B^*}}-\gamma_5 F_{\mu\nu}^{B^*}) \rangle \nonumber\\
 & & -\frac{m_{B^*0}^2}{4} \langle (X_{\bar B^* \mu} - \bar B^*_\mu \gamma_5)
(X_{B^*\mu}-\gamma_5 B^*_\mu) \rangle ,
\nonumber\\
{\cal L}_{B^*}^{\rm Yukawa} &=& -\frac{g_B^*}{4} \langle   
(X_{\bar B^*\mu}+\bar B^*_\mu \gamma_5)
(\sigma +i\gamma_5{\mib \tau}\cdot {\mib \pi} ) 
(X_{B^*\mu}-\gamma_5 B^*_\mu) \rangle ,
\label{eqLagB*}
\end{eqnarray}
where $F_{\mu\nu}$ is the field strength tensor of the corresponding vector 
or axial vector mesons. 
The $B^*$-meson and $X_{B^*}$-meson have a common 
bare-mass $m_{B^*0}$ in ${\cal L}_{B^*}^{\rm free}$ due to chiral symmetry. 
The spin-spin splitting is considered by  
$m_{B0}^*$ being different from $m_{B0}$ in Eq.(\ref{eqLagB}).

Similarly, for $D$ and $D^*$ meson systems, 
the chiral Lagrangian
is obtained from ${\cal L}_B$ and ${\cal L}_{B^*}$ 
simply by the substitution,
\begin{eqnarray}
{\cal L}_D &=& {\cal L}_B|_{(B,X_B,m_{B0},g_B) \to 
(\bar D,X_{\bar D},m_{D0},g_D)} \nonumber\\
{\cal L}_{D^*} &=& {\cal L}_{B^*}|_{(B^*,X_{B^*},m_{B^*0},g_B^*) \to 
(\bar D^*,X_{\bar D^*},m_{D^*0},g_D^*)} . 
\end{eqnarray}

The total Lagrangian ${\cal L}^{\rm tot}$ for ground-state 
$B$ and $D$ systems is given by
\begin{eqnarray}
{\cal L}^{\rm tot} &=& {\cal L}_B+{\cal L}_D+{\cal L}_{B^*}+{\cal L}_{D^*}
+{\cal L}_{L\sigma M},   
\label{eqtot} 
\end{eqnarray}
where ${\cal L}_{L \sigma M}$ is the Lagrangian of SU(2) L$\sigma$M,\cite{rfmw}
describing the spontaneous chiral symmetry breaking. Apparently,
Eq.~(\ref{eqtot}) is possibly the most simple Lagrangian with interaction
satisfying chiral symmetry.

Before going further,  we consider the constraints
on the present Lagrangian Eq.(\ref{eqtot}) required from HQS. 
The heavy quark spin symmetry and flavor symmetry predict, 
in this particular model,
respectively, the relations among the relevant one-pion emission amplitudes,
\begin{eqnarray}
&-g_B& \sqrt{\frac{1}{(2\pi )^9}} \sqrt{\frac{1}{ 2m_{X_B}  2m_B  2E_\pi }  }    
 = -g_B^* \sqrt{\frac{1}{(2\pi )^9}} \sqrt{\frac{1}{ 2m_{X_{B^*}} 2m_{B^*} 2E_\pi }} \nonumber\\
&=& -g_D \sqrt{ \frac{1}{(2\pi )^9}} \sqrt{\frac{1}{ 2m_{X_D} 2m_D 2E_\pi }}    
 = -g_D^* \sqrt{\frac{1}{(2\pi )^9}} \sqrt{\frac{1}{ 2m_{X_{D^*}} 2m_{D^*} 2E_\pi }} \nonumber\\   
&\equiv&  
- \eta  \frac{1}{(2\pi )^3 2} \sqrt{\frac{1}{(2\pi )^3 2E_\pi }}    ,
\label{eqeta}
\end{eqnarray}
where the universal amplitude $\eta$ in the present scheme 
corresponds to the amplitude  
$\xi$ defined in the general case (Eq.(\ref{eqxi})).

\section{Properties of Scalar and Axial-Vector Mesons}

In the mechanism of spontaneous chiral symmetry breaking the $\sigma$ 
acquires vacuum expectation value 
$\langle \sigma \rangle_0 \equiv \sigma_0 = f_\pi$,
which is described by ${\cal L}_{L\sigma M}$ in Eq.(\ref{eqtot}). 
Through the ${\cal L}^{\rm Yukawa}$
the mass degeneracies between scalar and pseudoscalar and between vector
and axial-vector are broken as
\begin{eqnarray}
m_{X_B}^2 &=& m_{B0}^2 + g_B f_\pi ,\ \ m_B^2 = m_{B0}^2 - g_B f_\pi ,\nonumber\\
m_{X_D}^2 &=& m_{D0}^2 + g_D f_\pi ,\ \ m_D^2 = m_{D0}^2 - g_D f_\pi ,\nonumber\\
m_{X_{B^*}}^2 &=& m_{B^*0}^2 + g_B^* f_\pi ,\ \ 
m_{B^*}^2 = m_{B^*0}^2 - g_B^* f_\pi ,\nonumber\\
m_{X_{D^*}}^2 &=& m_{D^*0}^2 + g_D^* f_\pi ,\ \ 
m_{D^*}^2 = m_{D^*0}^2 - g_D^* f_\pi ,
\label{eqmass}
\end{eqnarray}
which leads to 
\begin{eqnarray}
\Delta m_B &=& \Delta m_{B^*} = \Delta m_D = \Delta m_{D^*} \equiv \Delta m ,
\label{eqm}
\end{eqnarray}
because of Eq.(\ref{eqeta}), 
where the approximations $m_{X_B}\approx m_B \approx m_{B0}$ etc. are used.
Due to Eq.(\ref{eqhqrel}),
Eq.(\ref{eqm}) leads to the universality of decay widths,
\begin{eqnarray}
\Gamma_{X_B} &=&
\Gamma_{X_D}=\Gamma_{X_{B^*}} = \Gamma_{X_{D^*}}
=\Gamma (\Delta m).
\label{eqw}
\end{eqnarray}

In this model the $\Delta m$ is given by
\begin{eqnarray}
\Delta m &=& \eta f_\pi  ,
\end{eqnarray}
and thus, the $\Gamma (\Delta m)$ is given 
by Eq.(\ref{eqhqrel})
\begin{eqnarray}
\Gamma (\Delta m) &=&
 = 3 \frac{\sqrt{(\Delta m)^2 -m_\pi^2}}{8\pi} \xi^2 ,\ {\rm with}\ \ 
\xi =\eta = \Delta m / f_\pi . 
\label{eqUG} 
\end{eqnarray} 
The numerical value of $\Gamma (\Delta m)$ (with $f_\pi =93$MeV)
is depicted by solid line in Fig.~1.

The universality of mass splittings between the relevant chiral partners
and of the decay widths, which was derived by using the Yukawa coupling
in the most simple form, Eqs.(\ref{eqLagB}) and (\ref{eqLagB*}),
is shown to be rather generally valid from the following considerations. 
The mass-splittings and decay widths can be  
described more generally in the present scheme
of spontaneous chiral symmetry breaking  
by the following Lagrangian,
\begin{eqnarray}
{\cal L}_n^{\rm Yukawa} &=& -\sum_{M=B,\bar D}
 \frac{g_{M,n}}{4}\langle 
(X_{\bar M}-\bar M i\gamma_5)
(\sigma +i\gamma_5{\mib \tau}\cdot {\mib \pi} )
(\sigma^2+{\mib \pi}^2)^n 
(X_M-i\gamma_5 M) \rangle \nonumber\\
-  \sum_{M=B^*,\bar D^*}  & \frac{g_{M,n}}{4} &  \langle 
(X_{\bar M\mu} + \bar M_\mu \gamma_5)
(\sigma +i\gamma_5{\mib \tau}\cdot {\mib \pi})(\sigma^2+{\mib \pi}^2)^n 
(X_{M\mu} -\gamma_5 M_\mu) \rangle \ \ \ \ \ \ \ \ \ 
\label{eqYn}\\
{\cal L}_n^{\rm Z}\ \ \ \ \ \   &=& \sum_{M=B,\bar D} 
 \frac{h_{M,n}}{4}\langle 
\partial_\mu (X_{\bar M}-\bar M i\gamma_5) 
(\sigma +i\gamma_5{\mib \tau}\cdot {\mib \pi})(\sigma^2+{\mib \pi}^2)^n
 \partial_\mu (X_M-i\gamma_5 M) \rangle \nonumber \\
+  \sum_{M=B^* ,\bar D^*} & \frac{h_{M,n}}{4} &  \langle 
\partial_\mu (X_{\bar M\nu} +\bar M_\nu  \gamma_5) 
(\sigma +i\gamma_5{\mib \tau}\cdot {\mib \pi} ) 
(\sigma^2+{\mib \pi}^2)^n
 \partial_\mu (X_{M\nu} -\gamma_5 M_\nu) \rangle \ \ \ \ \ \ \ \ \ 
\label{eqZn}
\end{eqnarray}
with $n=0,1,2,...$ . 
When the $\sigma$ takes VEV, $\sigma_0=f_\pi$, the effects from the terms 
with $n\geq 2$ on the mass splittings and on the one pion emission amplitude 
can be included in the $n=0$ term through the redefinition, 
\begin{eqnarray}
g_{M,0} &\to & g_M \equiv 
\sum_{n=0,1,2,...} g_{M,n}f_\pi^{2n}
\ \ {\rm and}\ \ 
h_{M,0} \to  h_M \equiv
\sum_{n=0,1,2,...} h_{M,n}f_\pi^{2n} .\ \ \ 
\end{eqnarray}
The modified $n=0$ term of Eq.(\ref{eqYn}) is the simplest Yukawa Lagrangian, 
given in Eqs. (\ref{eqLagB}) and (\ref{eqLagB*}),
leading to the mass formula, Eq.(\ref{eqmass}), and to the  
one pion emission amplitude, Eq.(\ref{eqeta}). 
The modified $n=0$ term of Eq.(\ref{eqZn}) leads to the field renormalization,
\begin{eqnarray}
\sqrt{1-h_Mf_\pi}X_M &\equiv& X_M^{(r)},\ \  \sqrt{1+h_Mf_\pi}M\equiv M^{(r)},
\ \ (M=B,D,B^*,D^*),\ \ \ \ 
\end{eqnarray}
which gives, in the case of $h_Mf_\pi\ll 1$,
the same mass formula as Eq.(\ref{eqmass}) with $g_M$  
replaced by $h_M m_{M0}^2$.
This term also gives the one pion emission amplitude
\begin{eqnarray} 
&h_M & P_{X_M}\cdot P_M \sqrt{\frac{1}{(2\pi )^9}} \sqrt{\frac{1}{2E_{X_M} 2E_M 2E_\pi }}
\approx -h_M m_{X_M}m_M \sqrt{\frac{1}{(2\pi )^9}} \sqrt{\frac{1}{2m_{X_M} 2m_M 2E_\pi }}
\nonumber\\
&\approx& -h_M m_{M0}^2 \sqrt{\frac{1}{(2\pi )^9}} \sqrt{\frac{1}{2m_{X_M} 2m_M 2E_\pi }} ,
\end{eqnarray}
which is given from the LHS of Eq.(\ref{eqeta}) 
by replacing $g_M$ with $h_M m_{M0}^2$.

\begin{figure}[t]
  \epsfysize=8. cm
  \centerline{\epsffile{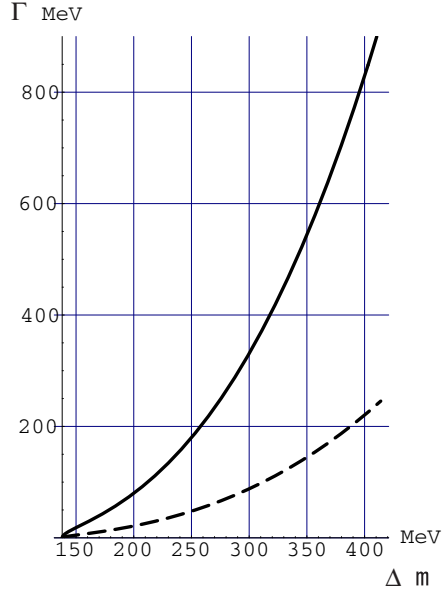}}
  \caption{$\Delta m$-dependence of the universal one-pion-emission decay width 
             $\Gamma (\Delta m)$ of $X_B,X_D,X_{B^*}$ and $X_{D^*}$ mesons:
             The $\Delta m = m_{X_M}-m_M\ (M=B,D,B^*,D^*)$ and 
             $\Gamma (\Delta m)$ take a common value due to HQS and chiral symmetry.
             The solid line represents the prediction by using the Lagrangian 
             in Eqs.(\ref{eqLagB}) and (\ref{eqLagB*}).
             The dashed line represents the prediction in the case of including
                      the correction of Eq.(\ref{eqLagd}) with the  
             $k=1/m_q^{\rm constit}=2/m_\rho $.  }
  \label{fig1}
\end{figure}

Thus, the effects from Eqs.(\ref{eqYn}) and (\ref{eqZn}) can be  
described by the simplest Yukawa Lagrangian in Eqs. (\ref{eqLagB}) 
and (\ref{eqLagB*}). 
Thus, its result of the universality of mass-splittings Eq.(\ref{eqm}), 
and of the universality of decay widths Eq.(\ref{eqw}) is kept
to be valid  
even in the case of taking the general contributions from Eqs. (\ref{eqYn}) 
and (\ref{eqZn}) into account.
However, the value of universal $\Gamma (\Delta m)$ itself generally becomes 
different from
the original one, Eq.(\ref{eqUG}),
by considering the effect from the following type of Lagrangian,
\begin{eqnarray}
{\cal L}^{\rm (d)} &=& \sum_{M=B,\bar D}\frac{k_M}{4}\langle 
\{    (X_M-i\gamma_5 M) \stackrel{\leftrightarrow}{\partial_\mu}
(X_{\bar M}-\bar M i\gamma_5)  \}
\partial_\mu (\sigma +i\gamma_5{\mib \tau}\cdot {\mib \pi})
 \rangle \nonumber\\
&+&  \sum_{M=B^* ,\bar D^*} \frac{k_M}{4}\langle 
\{ (X_{M\nu} -\gamma_5 M_\nu) \stackrel{\leftrightarrow}{\partial_\mu}
(X_{\bar M\nu} +\bar M_\nu  \gamma_5)  \}
\partial_\mu (\sigma +i\gamma_5{\mib \tau}\cdot {\mib \pi} ) 
 \rangle ,\ \ \ \ \ \ \ \ \ \ \ \ \ 
\label{eqLagd}
\end{eqnarray}
where $k_B= k_B^* = k_D = k_D^* \equiv k$ due to HQS.
By taking this Lagrangian into account, 
the Lorentz invariant amplitude of one pion emission decay
$-g_M (= -\eta \sqrt{m_{X_M} m_M}  \approx -\eta\ m_{X_M}
=-\Delta m \  m_{X_M}/f_\pi $ due to Eq.(\ref{eqeta}) ) 
is changed as  
\begin{eqnarray}
-g_M & \to &  -g_M - k\ p_\pi \cdot (P_M+P_{X_M}) \approx
 -g_M + 2 k \Delta m \ m_{X_M} \nonumber \\
&=& -\Delta m \ m_{X_M}(\frac{1}{f_\pi} -2k),
\end{eqnarray}
and the $\Gamma (\Delta m)$ is still given by Eq.(\ref{eqUG}) 
with the new $\xi$
\begin{eqnarray}
\xi  &=& \Delta m (1/f_\pi -2k).
\end{eqnarray} 
The order of the magnitude of  
$k$ is inverse momentum of the light quark.
The $\Gamma (\Delta m)$ is given in the case of $k=1/m_q^{\rm constit}$(, 
where the constituent quark mass $m_q^{\rm constit}\approx m_\rho /2$), 
as an example, by dashed line in Fig.~1. 
As can be seen in this figure 
the absolute value of $\Gamma$ is largely dependent upon the value of $k$, 
and cannot be predicted in the present model.

Finally we consider the possible effects of mixing
between pseudo-scalar and axial-vector and between scalar and vector mesons.
The corresponding Lagrangian is given by
\begin{eqnarray}
{\cal L}^{\rm mix} &=& \sum_{M=B,\bar D} \frac{l_M}{4} \langle
(X_{\bar M^*\mu}+\bar M^*_\mu \gamma_5)i\gamma_5
(\partial_\mu X_M+i\gamma_5\partial_\mu M)  \nonumber\\
 && \ \ \ \ \ \ \ \ \ \ \ \ +(\partial_\mu X_{\bar M} +\partial_\mu \bar M i\gamma_5 )i\gamma_5
(X_{M^*\mu}-\gamma_5 M^*_\mu )  \rangle  \nonumber\\
&=& \sum_M  l_M [ -(X_{\bar M^*\mu}  \partial_\mu M 
+\partial_\mu \bar M X_{\bar M^*\mu} ) + i ( \bar M^*_\mu \partial_\mu X_M
- \partial_\mu X_{\bar M} M^*_\mu )  ] ,\ \ \ \ \ \ \ \ \ \ \ \ \ \ 
\label{eqLagmix}
\end{eqnarray}
where $l_B=l_D\equiv l$ due to HQS, and the magnitude of this $l$ is 
of the order of light quark momentum $\sim m_q^{\rm constit}$.
The equation (\ref{eqLagmix}) leads to the redefinition of the 
vector and axial-vector meson fields,
\begin{eqnarray}
{\mib X}_{\bar M^*\mu} &\equiv& X_{\bar M^*\mu} 
+ \frac{l}{m_{X_{M^*}}^2} \partial_\mu  M,\ \ 
{\mib X}_{M^*\mu} \equiv X_{M^*\mu} 
+ \frac{l}{m_{X_{M^*}}^2} \partial_\mu  M, \nonumber\\
\bar{\mib M}^*_\mu &\equiv& \bar M^*_\mu 
+ \frac{il}{m_{M^*}^2} \partial_\mu X_{\bar M},\ \ 
{\mib M}^*_\mu \equiv M^*_\mu - \frac{il}{m_{M^*}^2} \partial_\mu X_M .
\label{eqmix}
\end{eqnarray}
Correspondingly, the field renormalization for the 
scalar and pseudoscalar mesons is necessary.
\begin{eqnarray}
\sqrt{1-\frac{l^2}{m_{X_{M^*}}^2} } M & \to & M,\ \     
\sqrt{1-\frac{l^2}{m_{M^*}^2} } X_M  \to  X_M.     
\label{eqmix1}
\end{eqnarray}
The field renormalization changes the masses of scalar and 
pseudoscalar mesons commonly as
$m_M^2 \to m_M^2(1+l^2/m_M^2)$ and $m_{X_M}^2 
\to m_{X_M}^2(1+l^2/m_{X_M}^2)$. 
This correction factor is of order $(m_q/m_Q)^2$, 
and thus expected to be small. 
Furthermore, it does not affect the mass splitting $\Delta m_M$.
Thus, it is not necessary to take these mixing effects into account.

\section{Concluding Remarks}

We have investigated the properties of new scalar $X_B,X_D$ and axial vector 
$X_{B^*},X_{D^*}$ mesons,
which are chiral partners of the ground state pseudoscalar $B,D$ 
and vector $B^*,D^*$ mesons, respectively.

Through the mechanism of spontaneous chiral symmetry breaking 
the masses of chiral multiplets split universally, 
and the decay widths for one pion emission of
$X_B$, $X_D$, $X_{B^*}$ and $X_{D^*}$ generally take a common value
in the present framework, due to chiral and heavy quark symmetry.

Possible existence of these new mesons 
and the validity of above predicted property 
should be investigated from experimental
data. Recently the spectra of $D^*\pi$ and $B\pi$ invariant mass from weak 
$Z^0$-boson decay were analyzed, and some evidences 
for possible existence of the new 
axial-vector $X_{D^*}$ and scalar $X_B$ are suggested, respectively.
The preliminary values of masses are
$m_{X_{D^*}}\approx 2300$MeV\cite{rfTo}
 and $m_{X_B}\sim 5520$MeV,\cite{rfMa}
corresponding to
$\Delta m_{D^*}\approx 290$MeV
and
$\Delta m_B\sim 240$MeV, respectively.
However, in order to derive any definite conclusion,
the data with higher statistics are necessary.

\acknowledgements
One of the authors (M. I.) is grateful deeply to 
Professor M. Oka and Professor A. Hosaka for 
useful discussions and continual encouragements.

\appendix

\section{Derivation of Chiral and HQ Spin Transformation Rule from\\
Covariant Spinor WF}
In this appendix we derive the chiral and HQ spin transformation rules 
for the heavy-light quark mesons which are, {\it up\ to\ the\ phase\ factor}, derived
from the covariant
bi-spinor WF given in our previous paper.\cite{rfCLC}
The (annihilation part of) WF for the light-heavy 
$q\bar Q$ mesons, 
with four velocity $v_\mu$ are 
\begin{eqnarray}
(U_B^{(+)},D_{X_B}^{(+)},U_{B^{*(s)}}^{(+)},D_{X_{B^*}^{(s)}}^{(+)})
=\frac{1}{2\sqrt 2}(i\gamma_5,1,i\epsilon^{(s)}\cdot\gamma ,
 i\gamma_5\epsilon^{(s)}\cdot\gamma )(1+iv\cdot\gamma ),\ \ \ \ \ 
\end{eqnarray} 
respectively. (The index $(+)$ is omitted in the following.)
Corresponging to the ``$i\gamma_5$"-transformation 
for the light-quark component WF,
$u({\mib p},s)\to u'({\mib p},s)=i\gamma_5 u({\mib p},s)$, 
the above bi-spinor meson WF are transformed as
\begin{eqnarray} 
U_B &\to& i\gamma_5 U_B=-D_{X_B},
\ \  D_{X_B}\to i\gamma_5 D_{X_B}=U_B,\nonumber\\
U_{B^{*(s)}} &\to& i\gamma_5 U_{B^{*(s)}}=i D_{X_{B^*}^{(s)}},
\ \  D_{X_{B^*}^{(s)}} \to i\gamma_5 D_{X_{B^*}^{(s)}}=i U_{B^{*(+)}},
\end{eqnarray} 
which are represented 
by the $\chi$ field operation as 
\begin{eqnarray}
\chi a_{B({\mib p})}^\dagger | 0 \rangle &=&
 - a_{X_B({\mib p})}^\dagger | 0 \rangle ,
\ \ \chi a_{X_B({\mib p})}^\dagger | 0 \rangle 
   = a_{B({\mib p})}^\dagger | 0 \rangle ,\nonumber\\
\chi a_{B^*({\mib p},\epsilon^{(s)} )}^\dagger | 0 \rangle 
&=& i a_{X_{B^*}({\mib p},\epsilon^{(s)})}^\dagger | 0 \rangle ,
\ \  \chi a_{X_{B^*}({\mib p},\epsilon^{(s)} )}^\dagger | 0 \rangle 
= i a_{B^*({\mib p},\epsilon^{(s)})}^\dagger | 0 \rangle . 
\label{eqA3}
\end{eqnarray}
In terms of the local meson field operator, Eq.(\ref{eqA3})
is represented by
\begin{eqnarray} 
\chi (\bar B,X_{\bar B},\bar B^*,X_{\bar B^*}) \chi^\dagger 
=(-X_{\bar B},\bar B, i X_{\bar B^*}, i \bar B^*). 
\label{eqa4}
\end{eqnarray}
Through its hermitian conjugation 
(which converts the charge of the meson),
Eq.(\ref{eqa4}) leads to 
\begin{eqnarray}
\chi (B,X_B,B^*,X_{B^*}) \chi^\dagger =(-X_{B},B, -i X_{B^*}, -i B^*), 
\label{eqA5}
\end{eqnarray}
which is equivalent to
\begin{eqnarray}
\chi^\dagger (B,X_B,B^*,X_{B^*}) \chi =(X_{B},-B, i X_{B^*}, i B^*). 
\label{eq-5}
\end{eqnarray}
The $\chi^\dagger$ field operation corresponds to the 
``$-i\gamma_5$" transformation for the quark component WF, 
$u({\mib p},s)\to u'({\mib p},s)=-i\gamma_5u({\mib p},s)$.
The equation (\ref{eq-5}) is also derived in a different way.
The WF are given by
\begin{eqnarray}
\langle 0| \Phi\   a_{B({\mib p})}^\dagger | 0 \rangle ,\ 
\langle 0| \Phi\   a_{X_B({\mib p})}^\dagger | 0 \rangle ,\ 
\langle 0| \Phi\   a_{B^{*(s)}({\mib p})}^\dagger | 0 \rangle ,\ 
\langle 0| \Phi\   a_{X_{B^*}^{(s)}({\mib p})}^\dagger | 0 \rangle ,
\end{eqnarray}
where 
\begin{eqnarray}
\Phi &=& \int d{\mib p}' (a_{B({\mib p}')}U_B+ a_{X_B({\mib p}')}D_{X_B}
\nonumber\\
&& + \sum_s (a_{B^*({\mib p}',\epsilon^{(s)} )}U_{B^{*(s)}} 
+ a_{X_{B^*}({\mib p}',\epsilon^{(s)} )}D_{X_{B^*}^{(s)} } ) ).
\end{eqnarray} 
Under the $i\gamma_5$-transformation, 
$\langle 0| \Phi a_{B({\mib p})}^\dagger  | 0  \rangle \to i\gamma_5  
\langle 0| \Phi a_{B({\mib p})}^\dagger  | 0  \rangle  $, which corresponds to 
$\langle 0| \Phi a_{B({\mib p})}^\dagger  | 0  \rangle \to  
\langle 0| \Phi \chi a_{B({\mib p})}^\dagger  | 0  \rangle 
=\langle 0| \chi \  \chi^\dagger \Phi \chi \  a_{B({\mib p})}^\dagger  | 0  \rangle  $.
Thus, 
the transformation rule is given by 
$\chi^\dagger \Phi \chi = i\gamma_5 \Phi$,
which is equivalent to  
\begin{eqnarray}  
\chi^\dagger (B i\gamma_5 +X_B 1
&+& B^* i\epsilon\cdot\gamma + X_{B^*} i\gamma_5\epsilon\cdot\gamma )
\chi\nonumber\\ 
=i\gamma_5 (B i\gamma_5 &+& X_B 1
+ B^* i\epsilon\cdot\gamma + X_{B^*} i\gamma_5\epsilon\cdot\gamma ).
\label{eqA9}
\end{eqnarray}
The equation (\ref{eqA9}) is more simplified as 
\begin{eqnarray}
\chi^\dagger (B i\gamma_5 +X_B 1 + B^* \gamma_5 + X_{B^*} )\chi  
=i\gamma_5 (B i\gamma_5 +X_B 1 + B^* \gamma_5 + X_{B^*} ), 
\label{eqphi}
\end{eqnarray}
which is equivalent to Eq.(\ref{eq-5}).

The $\chi^\dagger$ transformation rule for charge conjugated
$b\bar n$ systems is given 
from the Pauli conjugate, $\Phi \to \gamma_4 \Phi^\dagger \gamma_4$, 
of Eq.(\ref{eqphi}) as
\begin{eqnarray}
\chi^\dagger (\bar B i\gamma_5 +X_{\bar B} 1
- \bar B^* \gamma_5 + X_{\bar B^*} )\chi 
=(\bar B i\gamma_5 +X_{\bar B} 1
- \bar B^* \gamma_5 + X_{\bar B^*} )i\gamma_5 .\ \ \ \ \ 
\label{eqpauli}
\end{eqnarray}
By using Eqs.(\ref{eqphi}) and (\ref{eqpauli})
we can derive the chiral transformation rules given in \S 3(, which 
are the $\chi^\dagger$ transformation,) of the text.

Next we consider the operation of $S_Q^z$ on our bi-spinor WF.  
We take the meson rest frame and the $z$-direction as 
the spin quantization axis. 
Representing the spin configurations as $|s_q^z,S_{\bar Q}^z \rangle$
( $s_q^z (S_{\bar Q}^z)$ denoting the third component of light quark 
(heavy anti-quark) spin), the bi-spinor WF, 
$U_B,U_{B^{*(0)}},D_{X_B}$ and $D_{X_{B^{*(0)}}}$, are given
symbolically by  
\begin{eqnarray}
U_B &=& \frac{1}{i\sqrt{2}}(|+-\rangle - |-+\rangle ),\ \  
U_{B^{*(0)}}=\frac{1}{\sqrt 2}|+-\rangle + |-+\rangle ,\nonumber\\
D_{X_B}&=& \frac{1}{\sqrt 2}|+-\rangle - |-+\rangle ,\ \   
D_{X_{B^*}^{(0)}}=-\frac{1}{\sqrt 2}(|+-\rangle + |-+\rangle ),
\end{eqnarray}
where in $U$-type bi-spinor WF both the light-quark and heavy-antiquark spinors are in positive energy state,
but in $D$-type bi-spinor WF the light-quark spinor is in ``negative energy" state.
The operation of $S_Q^z$ on WF 
is the multiplication of $-\sigma_3^*$ to the $\bar Q$-index of  
WF from left side (or equivalent to the multiplication of 
$-\sigma_3^\dagger=-\sigma_3$ from right side).
\begin{eqnarray}
U_{B({\mib 0})}(-\sigma_3) &=& i U_{B^{*(0)}({\mib 0})},\ \  
U_{B^{*(0)}({\mib 0})}(-\sigma_3)=-i U_{B({\mib 0})},\nonumber\\
D_{X_B({\mib 0})}(-\sigma_3) &=& D_{X_{B^*}^{(0)}({\mib 0})},\ \  
D_{X_{B^*}^{(0)}({\mib 0})}(-\sigma_3)=D_{X_B({\mib 0})} . 
\label{eqA13}
\end{eqnarray}
The equation (\ref{eqA13}) corresponds to the following relation for  
meson creation operator
\begin{eqnarray}
[2S_Q^z,a_{B({\mib 0})}^\dagger ] &=& i a_{B^{*(0)}({\mib 0})}^\dagger ,\  \  
[2S_Q^z,a_{B^{*(0)}({\mib 0})}^\dagger ]= -i a_{B({\mib 0})}^\dagger ,\nonumber\\ 
\ [   2S_Q^z,a_{ X_B({\mib 0}) }^\dagger ] &=& a_{ X_{B^*}^{(0)}({\mib 0}) }^\dagger ,\  \    
[2S_Q^z,a_{ X_{B^*}^{(0)}({\mib 0}) }^\dagger ] = a_{ X_B({\mib 0}) }^\dagger .
\label{eqA14}
\end{eqnarray}
The hermitian conjugate of the above equations leads to 
\begin{eqnarray}
[2S_Q^z,a_{B({\mib 0})}] &=& i a_{B^{*(0)}({\mib 0})},\ \ 
[2S_Q^z,a_{B^{*(0)}({\mib 0})}]= -i a_{B({\mib 0})},\nonumber\\
\ [  2S_Q^z,a_{X_B({\mib 0})}] &=& - a_{X_{B^*}^{(0)}({\mib 0})},\ \ 
[2S_Q^z,a_{X_{B^*}^{(0)}({\mib 0})} ]= - a_{X_B({\mib 0})}.
\label{eqA15}
\end{eqnarray}
The equations (\ref{eqA14}) and (\ref{eqA15}) are represented 
by using meson field operator symbolically as
\begin{eqnarray}
[2S_Q^z, \bar B ] &=& i \bar B^{*(0)},\ \  [2S_Q^z,\bar B^{*(0)} ]= -i \bar B,\nonumber\\
\ [   2S_Q^z,X_{\bar B} ] &=& X_{\bar B^*}^{(0)},\ \  
[2S_Q^z,X_{\bar B^*}^{(0)} ]=X_{\bar B};\ \ {\rm and}\nonumber\\ 
\ [   2S_Q^z, B ] &=& i  B^{*(0)},\ \  [2S_Q^z, B^{*(0)} ]= -i  B,\nonumber\\
\ [   2S_Q^z,X_B ] &=& -X_{B^*}^{(0)},\ \  
[2S_Q^z,X_{B^*}^{(0)} ]=-X_B,
\end{eqnarray} 
where only the components with velocity ${\mib v}={\mib 0}$ are 
selected in all fields. 
The above equations are useful to make the Lagrangian satisfying 
the heavy quark spin symmetry.

\end{document}